\documentstyle[preprint,aps]{revtex}

\input epsf

\newcommand{\nc}{\newcommand}
\nc{\sh}[1]{\!\not\! #1}
\nc{\ga}[1]{\gamma_{#1}}
\nc{\g}{\gamma}
\nc{\gm}{\gamma_{\mu}}
\nc{\gf}{\gamma_{5}}
\nc{\Tr}{ \, {\mbox {\rm Tr}}[}
\nc{\decay}[2]{\Gamma_{#1 \rightarrow #2}}
\nc{\lam}[1]{\lambda^{#1}(m_{1}^{2},m_{2}^{2},m_{3}^{2})}
\nc{\mod}[1]{|#1|}
\nc{\dg}[1]{#1^{\circ}}
\nc{\gvpg}{g_{_{VP}\g}}
\nc{\tV}{\theta_{V}}
\nc{\tP}{\theta_{P}}
\nc{\be}{\beta}

\nc{\dfp}{\frac{d^{4} p}{(2\pi)^4}}
\nc{\mb}{\bar{m}}
\nc{\Qb}{\bar{Q}}
\nc{\gP}{g_{Pq\bar{q}'}}
\nc{\gV}{g_{Vq\bar{q}'}}
\nc{\gVu}{g_{Vu\bar{u}}}
\nc{\gPu}{g_{Pu\bar{u}}}
\nc{\gVs}{g_{Vs\bar{s}}}
\nc{\gPs}{g_{Ps\bar{s}}}  
\nc{\gVc}{g_{Vc\bar{c}}}
\nc{\gPc}{g_{Pc\bar{c}}} 
\nc{\Jo}{J_1}
\nc{\Jt}{J_2}
\nc{\Jb}{\bar{J}}
\nc{\Jcl}{J^{{\rm CL}}}
\nc{\Jhql}{J^{{\rm HQL}}}
\nc{\Jbhql}{\bar{J}^{{\rm HQL}}}
\nc{\dil}[1]{\mbox{Li}_{2}(#1)}
\nc{\vot}{v_{1,2}}

\begin{document}
\preprint{UTAS-PHYS-95-48}
\draft

\title{Radiative decays of heavy and light mesons \\ in a quark triangle 
approach}

\author{N.R. Jones and Dongsheng Liu}

\address{
Physics Department, University of Tasmania,\\
GPO Box 252C, Hobart, Tasmania 7001, Australia.}

\maketitle

\begin{abstract}
The radiative meson decays $V\to P\gamma$ and $P\to \gamma\gamma$ are analyzed using
the quark triangle diagram.  Experimental data yield well determined
estimates of the universal quark-antiquark-meson couplings $\gV$ and $\gP$
for the light meson sector.  Also predictions for the ratios of neutral to
charged heavy meson decay coupling constants are given and await
experimental confirmation. 
\end{abstract}


\section{Introduction}
In earlier work \cite{jones94} we used a supermultiplet theory uniting the
vector and pseudoscalar mesons to attempt to obtain a universal
three--point coupling constant. The relativistic multispinor fields of the
supermultiplet theory described pointlike mesons with correct spin,
parity, flavor and color degrees of freedom without necessarily invoking
the notion of constituent quarks.  Despite the apparent contradiction with
the modern understanding of the quark nature of mesons it has been shown
\cite{huss91b} that such a field is dynamically equivalent to a system of
two quarks moving at equal velocity and on--shell.  This relatively simple
scheme does nonetheless compare favorably with experimental results for
strong vector and pseudoscalar interactions in the light and heavy meson
sector. When we include the radiative decays by incorporating vector meson
dominance with the scheme we once again found reasonable agreement, but
there were some unexpected and significant deviations from the theory,
particularly in the $K^*\to K\gamma$ decays. 

The supermultiplet theory and exact SU(3) predict the coupling ratio
$\mod{ g_{K^{*0} K^0 \gamma}/g_{K^{*+} K^+ \gamma}}$ to equal 2.  But experimental
measurement currently estimates the ratio as $1.51 \pm 0.13$, a
substantial difference. One possible reason why the supermultiplet scheme
did not comply with the experimental measure is that the exact form of the
vector meson dominance is not known in the $q^2 \to 0$ limit; it is only
accurately known for $q^2 = m_V^2$ from $V \to l^+l^-$ decays.  Hence
there is some uncertainty in extrapolating vector meson dominance to the
off--shell case. However, the difference in theory and experiment is so
large that it is unlikely this is the only contributing factor. Thus we
also implemented some symmetry breaking in the supermultiplet scheme, but
the $K^*$ radiative decays seemed impervious to our attempts at matching
theory with experiment as large discrepancies remained. 

To understand these deviations further, we sought a method which easily
allowed for off--shell propagation of the quarks so that we could evaluate
the magnitude of this necessary correction. A convenient and apparently
successful method for doing this was by use of a quark triangle diagram;
which so far has given accurate predictions for $\pi^0 \to \gamma\gamma$ decay
widths \cite{amet84} and pion and kaon charge radii \cite{amet84,tarr}.  A
form which used chiral and isospin symmetry has also been successfully
applied to the $K^*$ radiative decay problem and some radiative decays in
the light meson sector \cite{bram8912}.  The resulting loop integral
accounted for the difference in quark masses and hence propagators in the
loop, and correspondence between theory and experiment was achieved. 

Nonetheless, a crucial assumption of the quark triangle diagram is that
the meson--quark--antiquark vertex has the form $\gP \gamma_5$ for the
pseudoscalar meson and $\gV \gamma^{\mu}$ for the vector meson. If we are
to confidently use the quark triangle method we wish to test the
appropriateness of this assumption.  We do this by extracting the
couplings from experimental measurements and examine the extent to which
they carry the spin and flavor symmetries. With this in mind, we firstly
formulate the limit free form of the integral and derive a limiting case
which uses chiral symmetry.  This enables us to compare our result with
others.  We also determine a heavy quark expansion.  Following this, the
scheme is applied to $P\to \gamma\gamma$ decays which yield estimates of $\gP$ and
then to $V\to P\gamma$ decays to obtain the product $\gV\gP$ for different
quark flavors. 

The results indicate that the meson--quark--antiquark couplings in the
$VVP$ sector determined from different channels which involve common
constituent quarks are remarkably uniform, suggesting that the effective
vertex in the quark triangle diagram is valid. The data also demonstrates
that the triangle method should be highly predictive due to the stability
of the couplings. Finally we use the method in the heavy meson sector, to
predict coupling ratios of the form $g_{V^0 P^0 \gamma}/g_{V^+ P^+ \gamma}$ where
the only free parameters required are the constituent quark masses. Our
result for $D^*$ decays fall within other theoretical estimates, while
that of $B^*$ is sensitive to $b$ quark mass. 

\section{The Quark Triangle}
\begin{figure}
\begin{minipage}[h]{6.5cm}
\epsfxsize=7cm  \epsfbox{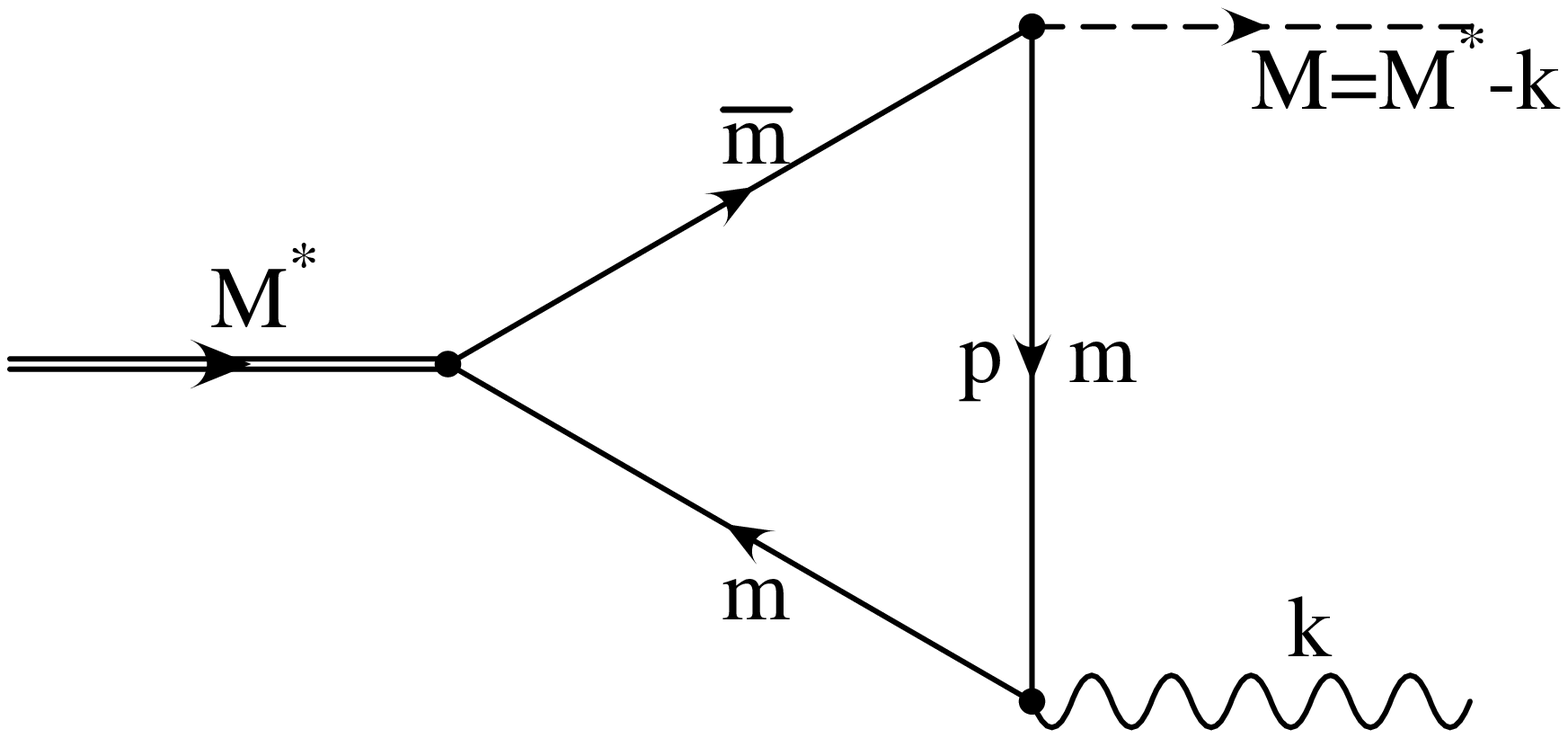}
\end{minipage}\ + \
\begin{minipage}[h]{6.5cm}
\epsfxsize=7cm \epsfbox{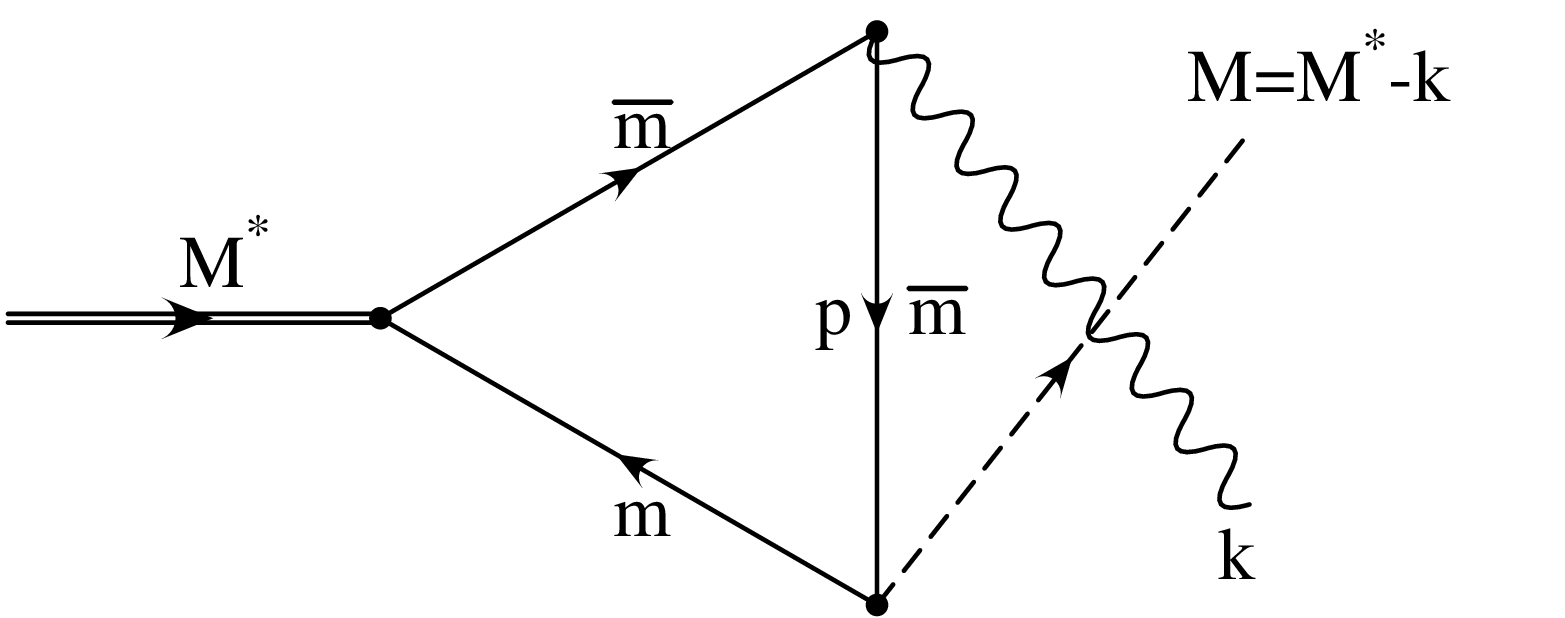}
\end{minipage}
\vspace*{3mm}
\caption{Quark triangle diagrams contributing to $V\to P\gamma$.}
\label{fig:qtri}
\end{figure}

In the quark triangle formulation of Figure~\ref{fig:qtri}, the decay
from vector meson to pseudoscalar meson and photon state is mediated
by a quark loop with flavors of constituent mass $m$ and $\bar{m}$.
(The choice of constituent mass rather than current mass is supported 
\cite{amet84,tarr}).
The quark triangle diagrams correspond to the Feynman amplitude for the
decay, 
\begin{eqnarray} 
A(V\to P\gamma) &=& -N_C \gV \gP e Q \kappa^{\mu} \epsilon^{\nu} \nonumber \\
&& \hspace*{-0.5cm} \int \dfp \Tr\gamma_{\nu} \frac{1}{\sh{p} -m} \gamma^5
\frac{1}{\sh{p} + \sh{M}^* - \sh{k} - \mb} \gamma_{\mu} \frac{1}{\sh{p} -
\sh{k} - m} ]  + 
(m \leftrightarrow \bar{m}, Q \leftrightarrow \bar{Q}), \label{eq:Amp}
\end{eqnarray}
where $\kappa^{\mu} (\epsilon^{\nu})$ is the vector meson (photon)
polarization vector, $M^* (M)$ is the vector (pseudoscalar)
four--momenta, $\gV (\gP)$ is the vector--quark (pseudoscalar--quark)
coupling constant and $e Q$ is the electric charge of the quark of
mass $m$ in the loop.

The Feynman loop integral involved in the amplitude (\ref{eq:Amp}) may
be solved with standard techniques.  We maintain the notation of
Bramon and Scadron \cite{bram8912} and call the integral $J$, a 
dimensionless quantity after multiplication by $m$. Subsequently,
\begin{eqnarray}
J &=& m \int_{0}^{1}du \int_{0}^{1-u} dv \frac{m + (\mb-m)u}{
 m^2 + (\mb^2-m^2) u - M^{*2} uv - M^2 u(1-u-v)} \\  \label{eq:Ji}
&=& -\,\frac{m^2}{M^{*2}-M^2} \int_{0}^{1}du (\delta + 1/u) ( \ln
\mod{j^*} - \ln \mod{j} ) \\ \nonumber
&=& \frac{m^2}{M^{*2}-M^2} [\Jo^* - \Jo + \Jt^* - \Jt], \nonumber
\end{eqnarray}
where 
\begin{eqnarray*}
 \delta &=& \frac{m-\bar{m}}{m} \\
 j &=& [m^2-(m^2-\mb^2+M^2)u+M^2 u^2]/m^2 \\
 \Jo &=& -\int_{0}^{1}du \ln\mod{j}/u \\
 \Jt &=& -\delta \int_{0}^{1}du \ln\mod{j},
\end{eqnarray*}
and $j^* (\Jo^*, \Jt^*)$ corresponds to $j (\Jo, \Jt)$ with $M
\leftrightarrow M^*$, respectively.  Now $M^* (M)$ corresponds 
to the vector (pseudoscalar) mass.

\subsection{Determination of $\Jo$}
In attempting to find an expression for $\Jo$, we re--write the
argument of the natural log in a form similar to that of the
dilogarithm.   To do this, we factorize $j$ as
\[ 
j=1+(\delta^2+2\delta- (M/m)^2)u+M^2 u^2/m^2 = (1-v_1 u)(1-v_2 u),
\]
where
\begin{eqnarray}
 \vot &=& - \{\delta^2+2\delta- (M/m)^2 \mp [(\delta^2+2\delta- 
     (M/m)^2)^2 - \,(2 M/m)^2]^{1/2} \}/2 \label{eq:vot1} \\
 &=&  \{ m^2 - \mb^2 + M^2 \pm \lambda^{1/2}(m^2,\mb^2,M^2) \}/2m^2. 
	\label{eq:vot2}
\end{eqnarray}
and
\begin{equation}
 \lambda(m^2,\mb^2,M^2) = [M^2 - (m+\mb)^2][M^2 - (m-\mb)^2].
 \label{eq:Lambda}
\end{equation}
We obtain a similar expression for $\vot^*$ upon substitution of $M$
by $M^*$. The factorization we have performed does not necessarily
lead to real $v_k^*$ or $v_k$ and we must consider the case for both
real and complex arguments.

\subsubsection{Real $v_k^*$ or $v_k$}
We first consider the simplest case, that when either $v_k^*$ or
$v_k$ is real. For the moment we simply deal with real $v_k$ and extend 
our findings to real $v_k^*$ by substitution of $M$ by $M^*$.  
The factor $v_k$ is only real when $\lambda(m^2,\mb^2,M^2) \geq 0$ which
from (\ref{eq:Lambda}) implies
\begin{equation}
 M \geq  m + \mb \: \mbox{ or } \: M \leq \mod{m - \mb}.
	\label{eq:Mcondition}
\end{equation}
When we are assured of real $v_k$ the solution of $\Jo$
is related to the standard dilogarithm function,
\begin{equation}       
\Jo = - \int_0^1 du \sum_{k=1}^2 \frac{\ln \mod{1- v_k u}}{u} 
\equiv  \sum_{k=1}^2 \dil{v_k,0} = 
	\left\{ \begin{array}{l}
	\sum_{i=k}^2 \dil{v_k} \;\; \mbox{for} \;\; v_k \leq 1 \\
	\sum_{k=1}^2 \dil{v_k} + i \pi \ln \mod{v_k} 
	\;\; \mbox{for} \;\; v_k > 1,
	\end{array} \right. \label{eq:realJ1}
\end{equation}

\subsubsection{Complex $v_k^*$ or $v_k$}
The $v_k$ are complex if $\lambda(m^2,\mb^2,M^2) < 0$ and we
need an appropriate method for handling this situation.  Fortunately
the dilogarithm of a complex argument does exist so we may proceed.

We express $\Jo$ as
\begin{eqnarray}
 \Jo &=& - \int_0^1 du \sum_{k=1}^2 
	\frac{\ln \mod{1- v_k u}}{u} = 
	- \sum_{k=1}^2 \int_0^{\rho e^{i \phi_k}}
	\frac{\ln(1-z)}{z} dz \nonumber \\
 &=& - \sum_{k=1}^2\frac{1}{2} \int_0^{\rho} 
	\frac{\ln(1-2x\cos\phi_k+x^2)}{x}dx + 
	i\int_0^{\rho} {\rm arctan} \left[ \frac{y \sin\phi_k}{1-
	y\cos\phi_k}       \right] \frac{dy}{y} \nonumber \\
\Jo &=& -\int_0^{\rho} \frac{\ln(1-2x\cos\phi+x^2)}{x}dx \nonumber \\
 & \equiv & 2 \dil{\rho,\phi}, \label{eq:complexJ1}
\end{eqnarray}
where
\begin{equation} \rho = M/m, \;\; \phi = \phi_1 = -\phi_2 \;\; {\rm and} \;\;
 \cos \phi = \frac{m^2-\mb^2+M^2}{2Mm},
\end{equation}
and $\Jo^*$ corresponds to $\Jo$ with $M \leftrightarrow M^*$.

\subsection{Determination of $\Jt$}
$\Jt$ is a simpler integral to evaluate as it does not contain the $1/u$
dependence.  Recall,
\[ 
\Jt = -\delta \int_{0}^{1}du  \ln [1-(1-\mb^2/m^2+M^2/m^2)u+M^2 u^2/m^2],
\]
which, by standard techniques, reduces to
\[
 \Jt =  \frac{m-\mb}{m \; M^2} \left[ (\mb^2-m^2+M^2) \ln \frac{\mb}{m} 
	 - \lambda^{1/2}(m^2,\mb^2,M^2)
\mbox{arctanh}  \left( \frac{\lambda^{1/2}(m^2,\mb^2,M^2)}{\mb^2+m^2-M^2}
\right) \right],
\]
and is valid for all $m, \mb, M$. We obtain a similar expression
$\Jt^*$ when we substitute $M \to M^*$.

It is also useful to express $\Jt$ in terms of real $v_k$.  Following
a similar method to that used for deriving $\Jo$, we find
\begin{equation}
 \Jt = -\delta \sum_{k=1}^2 (1-1/v_k)\ln\mod{1-v_k} \; \mbox{ for } \; 
	\lambda(m^2,\mb^2,M^2) \geq 0.
\end{equation}

\section{Comparison with Covariant Amplitude}
Our final form for the loop integral is:
\begin{equation}
 J = \frac{m^2}{M^{*2}-M^2} [ \Jo^* - \Jo + \Jt^* - \Jt].
\label{eq:J}
\end{equation}
Here we have not considered the imaginary part in $\Jo$ which is 
irrelevant to the decay process, 
\begin{equation}
 \Jo = \left\{ \begin{array}{rl}
	\sum_{k=1}^{2} \dil{v_k,0} 
		 & \mbox{if  } \lambda(m^2,\mb^2,M^2) \geq 0 \\
	& \mbox{where  } \vot = [m^2-\mb^2+M^2 \pm 
			\lambda^{1/2}(m^2,\mb^2,M^2)]/2m^2  \\
	2 \dil{\rho,\phi}  
		& \mbox{if  } \lambda(m^2,\mb^2,M^2) < 0  \\
	&\mbox{where  } \rho = M/m, \; \cos \phi =  
		(m^2-\mb^2+M^2)/2Mm \\
	\end{array} \right.
\label{eq:J1}
\end{equation}
and
\[
 \Jt =  \frac{m-\mb}{m \; M^2} \left[ (\mb^2-m^2+M^2) \ln \frac{\mb}{m} 
	 - \lambda^{1/2}(m^2,\mb^2,M^2) 
\mbox{arctanh}  \left( \frac{\lambda^{1/2}(m^2,\mb^2,M^2)}{\mb^2+m^2-M^2}
\right) \right].
\]
Thus, our Feynman amplitude for the decay is
\begin{equation}
A(V\to P\gamma) = i N_C e \gV \gP  \epsilon_{\mu\nu\rho\sigma} \kappa^{\mu} 
	\epsilon^{\nu} P^{\rho} k^{\sigma} [ Q J /m + \Qb \Jb /\mb ]/4\pi^2,      
\label{eq:VPg}
\end{equation}
where $J \to \Jb$ when $m \leftrightarrow \mb$ (from the momentum crossed 
Feynman diagram) and $\Qb$ is the charge of the quark with mass 
$\mb$.  We compare this with the general covariant amplitude for the 
process $V \to P \gamma$ 
\[
 A(V \to P\gamma) = i \gvpg \epsilon_{\alpha\be\mu\nu} 
P^{\alpha} k^{\be} \kappa^{\mu} \epsilon^{\nu}
\]
so that our quark triangle approach resolves the $\gvpg$ covariant 
coupling constant as 
\begin{equation}
 \gvpg = N_C e \gV \gP [Q J /m + \Qb \Jb/\mb]/4\pi^2.
\label{eq:couple}
\end{equation}
 
\section{$P \to \gamma \gamma$ in the Quark Triangle Scheme}
We are interested in understanding the behaviour and obtaining actual
values for the coupling constants $\gP$ and $\gV$ in the light quark
sector so that we can use appropriate estimates for these couplings 
in the heavy quark sector. To this end we can use the well documented 
decay data for $V \to P \gamma$ in the light vector meson sector, as well 
as the decays $P \to \gamma\gamma$.  These latter processes are particularly 
useful as they only involve the coupling $\gP$, and not the product 
$\gV\gP$ as do the first case.  Consequently, we must derive the 
amplitude for the decay of a pseudoscalar meson into two photons.  
This we may do by following a similar derivation as above, but it is 
much simpler to make the following substitutions in the $V \to P \gamma$ 
amplitude (\ref{eq:VPg}): 
\[
 M^* \to M, \; M \to 0, \;  \gV \to \gP, \; 
\mbox{ and } \gP \to e Q,
\]
and since all pseudoscalar mesons involved in $P \to \gamma\gamma$ decays must
be quark flavor singlets, $m = \mb, \; Q = \bar{Q}, \; J = \Jb$.  
Subsequently (\ref{eq:couple}) is reduced to
\begin{equation}
 g_{P\gamma\gamma} = 2 N_C \gP e^2 [Q^2 J/m] /4\pi^2
\end{equation}
where 
\[
 J = \left\{ \begin{array}{ll}
	\frac{m^2}{M^2} \sum_{k=1}^{2} \dil{v_k,0} 
		 & \mbox{if  } M \geq 2m \\
	& \mbox{where  } \vot = M [M/m \pm (M^2/m^2-
				4)^{1/2}]/2m  \\
	2 \frac{m^2}{M^2} \dil{\rho,\phi}  
		& \mbox{if  }0\leq M\leq 2m \\
	&\mbox{where  } \rho = M/m, \cos \phi = M/2m \\
	\end{array} \right.
\]

\section{Chiral Limit}
The chiral symmetry limit is useful in the light meson sector, and gives 
our work direct comparison with that of Bramon and Scadron \cite{bram8912}.
The limit corresponds to a small pseudoscalar mass when
compared to the vector mass, that is $M^{*2} >> M^2$. Such a limit is
entirely appropriate for the study of the radiative decays of $K^*$ 
mesons, and it is re--assuring to know that our $J$ reduces to the
$J_M$ of \cite{bram8912} in the chiral limit.

The chiral limit, corresponding to $M \to 0$ in $J$, enables us to use
the real form for $\Jo$ in (\ref{eq:realJ1}) as $M \leq \mod{m-\mb}$.
\[ v_k(M \to 0) = \left\{ \begin{array}{l}
		0 \;\; \mbox{for} \;\; k=1 \\
		-(\delta^2 + 2 \delta) \;\; \mbox{for} \;\; k=2,
		 \end{array} \right.
\]
from (\ref{eq:vot1}) so that
\[ \Jo = \sum_{k=1}^2 \dil{v_k,0} = \dil{0,0} + \dil{-\delta^2-2\delta,0} = 
	\dil{-\delta^2-2\delta,0},
\]
and 
\begin{eqnarray*} \Jt &=& -\delta\sum_{k=1}^2 (1-1/v_k)\ln\mod{1-v_k} \\
 &=& - \delta - \delta[1+1/(\delta^2+2\delta)]\ln\mod{1+2\delta+\delta^2} =
	-\delta - \frac{2(\delta+1)^2}{\delta+2} \ln\mod{1+\delta}.
\end{eqnarray*}
Thus $J$ in the chiral limit, which we denote as $\Jcl$,
becomes
\begin{eqnarray}
 \Jcl &=& \frac{m^2}{M^{*2}} \left\{\delta + \sum_{k=1}^{2} [
\dil{v_k^*,0} - \delta(1-1/v_k^*)\ln\mod{1-v_k^*}] \ldots \right.\nonumber \\ 
 && \hspace*{3cm} \left. -\dil{-\delta^2-2\delta,0} +
\frac{2(1+\delta)^2}{2+\delta}\ln\mod{1+\delta} \right\},
\label{eq:Jcl}
\end{eqnarray}
where we have assumed $v_k^*$ is real.
This form may be simplified even further near the isospin symmetry
limit whereby $m \approx \mb$.  In this instance we ignore $\delta$
terms of order 2 and higher.  
\begin{eqnarray}
 \dil{-\delta^2-2\delta,0} &\rightarrow& \dil{-2\delta,0}, \nonumber \\ 
 \frac{2(1+\delta)^2}{2+\delta}\ln\mod{1+\delta} 
 &\rightarrow& (\delta+\frac{1}{2})\ln\mod{1+2\delta}, \nonumber \\
 \vot^* &\rightarrow& -\{ 2\delta- (M^*/m)^2 \mp [(2\delta+(M^*/m)^2)^2-
			(2M^*/m)^2]^{1/2} \}/2 \nonumber \\
 &=& \frac{M^{*2}}{2m^2} - \delta \pm \left[ \left(\frac{M^{*2}}{2m^2} - \delta
	 \right)^2 - \frac{M^{*2}}{m^2} \right]^{1/2}. \label{eq:viCIL}
\end{eqnarray}
Thus we find $\Jcl$ incorporating isospin symmetry between quark
flavors reduces to
\begin{eqnarray*} \Jcl &=& \frac{m^2}{M^{*2}} \left\{\delta + \sum_{k=1}^{2} [
	\dil{v_k^*,0} - \delta(1-1/v_k^*)\ln\mod{1-v_k^*}] \right. \\
&&      \left. -\dil{-2\delta,0} +
	(\delta+\frac{1}{2})\ln\mod{1+2\delta} \:\right\}, 
\end{eqnarray*}
with $v_k^*$ defined in Eq.\ (\ref{eq:viCIL}).  This form is very similar to
the $J_M$ of Bramon and Scadron \cite{bram8912}. There is a subtle
difference in their use of the dilogarithm function $\dil{z}$ versus our 
function $\dil{z,0}$ which is similar to the dilog, but which only allows 
real solutions (the term $i\pi\ln\mod{v_k}$ in Eq.\ (\ref{eq:realJ1}) 
ensures this).  In the chiral limit with $M^*>m+\mb$ these two
functions are equivalent.
In addition they have the term $(\delta - 1/2)\ln\mod{1+2\delta}$ whereas we have 
$(\delta+1/2)\ln\mod{1+2\delta}$.  We believe this difference is due to a 
typographical mistake as the argument of the log function is linked to 
the multiplier outside, so that there should be no difference between 
them (the missing multiplication factor of two is easily accounted for, 
but not the sign change).

\subsection{Chiral Limit of $P \to \gamma\gamma$}
There is a well-known chiral limit of the $P \to \gamma\gamma$ case, namely 
$\pi^0 \to \gamma\gamma$ \cite{bern,amet84}, when 
$g_{\pi^0\gamma\gamma} = e^2 N_C g_{\pi^0} Q^2/4 \pi^2 m $.  This implies that 
$J = 1/2$ for the pion.  We can establish this from 
our full formulae.  The chiral limit implies that $M_{\pi^0} \to 0$
so that the appropriate form of $J$ is
\[ J = 2 m^2 \dil{\rho,\phi} /M^2, \]
and we define $M/m = \epsilon$ with $\epsilon \to 0$ as $M \to 0$.  We
subsequently find $\rho = \epsilon, \;\; \cos \phi = \epsilon/2 $
and therefore
\begin{eqnarray*}
 J &=& \frac{2}{\epsilon^2} [ -\frac{1}{2} \int_{0}^{\epsilon} 
	\frac{ \ln(1-\epsilon x+x^2)}{x} dx ] \\
 &\approx& - \frac{1}{\epsilon^2} \int_{0}^{\epsilon} dx [-\epsilon + 
	(1-\epsilon^2/2)x+ \epsilon(\epsilon^2+3)x/3+ \ldots ] \\
 &=& \frac{1}{2} - \epsilon^2/12 + \epsilon^4/9 + \ldots
\end{eqnarray*}
Therefore $J$ reproduces the $\pi^0 \to \gamma\gamma$ result in the chiral limit.

\section{Heavy Quark Expansion}
Since we are particularly interested in the heavy meson decays $D^* \to 
D\gamma$ and $B^*\to B\gamma$ we feel it is of interest to examine the heavy 
quark expansion of our loop integral $J$. To derive this 
we consider an expansion in terms of the light 
to heavy quark mass ratio in each of the loop integrals.  We make
the arbitrary choice of $m=m_q$ and $\mb=m_Q$, where $m_q$ is the
light quark mass, and $m_Q$ the heavy quark mass.  These lead to
the following definitions:
\begin{eqnarray}
 \frac{m}{\mb} &=& \epsilon  \label{eq:ep} \\
 M &=& \mb + \Lambda,       \label{eq:M}  \\
 M^* &=& \mb + \Lambda^*    \label{eq:M*} 
\end{eqnarray}
where $\epsilon \to 0$ in the heavy quark limit, and $\Lambda, \; 
\Lambda^*$ is the combined binding energy and light quark mass.

\subsection{Heavy Quark Expansion of $J$}
When dealing with $J$, Eq.\ (\ref{eq:M}) and (\ref{eq:M*}) lead to 
\begin{eqnarray*}
 M/m =  r + 1/\epsilon, \\
 M^*/m = r^* + 1/\epsilon
\end{eqnarray*}
where $r=\Lambda/m$ and $r^*=\Lambda^*/m$ which are independent of the heavy 
quark mass.  We express $J$ of Eq.\ (\ref{eq:Ji}) using these relations 
to find
\[ 
J \approx \int_0^1 du \int_0^{1-u} dv 
\frac{u/\epsilon}{u/\epsilon^2-2(r^*-r)uv/\epsilon-(2r+1/\epsilon) 
u(1-u)/\epsilon} 
\]
ignoring the constant term in $\epsilon$.  This reduces to 
\[
 \Jhql =  - \, \frac{\epsilon (r^* \ln \mod{2 r^* \epsilon} - 
 r \ln\mod{2r \epsilon})}{ (r^*-r)(1+2\epsilon(r^*+r))}
\]
as the highest order terms in the expansion.  Note that this term is of 
the form $\epsilon\ln\epsilon$ and we can thus expect slow convergence 
of the heavy quark expansion.

\subsection{Heavy Quark Expansion of $\Jb$}
We maintain our definition of $\epsilon$ and $r$; however since $\Jb$
corresponds to $J$ with $m \leftrightarrow \mb$; we shall need
the relations
\begin{eqnarray*}
 M/\mb = 1 + \Lambda/\mb = 1 + r \epsilon \\
 M^*/\mb = 1 + \Lambda^*/\mb = 1 + r^* \epsilon,
\end{eqnarray*}
from which 
\[ \Jb \approx \int_0^1 du \int_0^{1-u} dv \frac{ 1+(\epsilon-1)u}{ 
1-u-2(r^*-r)\epsilon u v - (1+2r\epsilon)u(1-u)}, \]
ignoring the $\epsilon^2$ contribution. Thus our heavy quark 
expansion of $\Jb$ reduces to 
\[ \Jbhql = [\dil{1+2r^*\epsilon,0}-\dil{1+2r\epsilon,0}+
 \frac{2\epsilon(1-\epsilon)} {1+2\epsilon(r^*+r)} 
 (r^*\ln\mod{2r^*\epsilon}-r\ln\mod{2r\epsilon})]. \]
To simplify this form further, we consider an expansion of the 
dilogarithm.  Since
\begin{eqnarray*}
 \dil{1+2r^*\epsilon,0}-\dil{1+2r\epsilon,0} &=& 
	- \int_{1+2r\epsilon}^{1+2r^*\epsilon} du \frac{\ln\mod{1-u}}{u} \\
 &\approx& -\int_{2r\epsilon}^{2r^*\epsilon} dz \, \ln z (1-z+z^2-z^3+\ldots) \\
 &=& -2\epsilon ( r^*\ln\mod{2r^*\epsilon}-r\ln\mod{2r\epsilon}- 
 (r^*-r) + {\cal O}(\epsilon) ), 
\end{eqnarray*}
then we arrive at our final form
\[  \Jbhql = 
 1- \frac{\epsilon}{r^*-r}(r^*\ln\mod{2r^*\epsilon}-r\ln\mod{2r\epsilon}).
\]
Once again observe the $\epsilon\ln\epsilon$ dependence, indicative of 
slow convergence.

It appears that both expansions $\Jhql$ and $\Jbhql$ will only
converge slowly to their true counterparts $J$ and $\Jb$.  Thus,
unfortunately, they are not so useful approximations for either the $c$
or $b$ quark cases.  

There are other possible expansions we could consider, namely that of the
$P\to\gamma\gamma$ and $V \to P\gamma$ loop integrals where there is only one quark
flavor in the loop (such as $\eta_c\to\gamma\gamma$, $\eta_b \to \gamma\gamma$, $J/\psi
\to \eta_c \gamma$ or $\Upsilon \to \eta_b \gamma$) and consider some expansion as
the quark mass becomes large.  Unfortunately, such an expansion fails to
be a good approximation, simply due to the assumption one has to make
about the pseudoscalar and/or vector mass.  For example, in the $P\to\gamma\gamma$
case one would assume the pseudoscalar mass $M$ would consist of the sum
of the quark masses along with some binding energy so that $M=2m+\Delta$.
Then the loop integral could be expressed as \[ J = \Jb = \int_0^1 du
\int_0^{1-u} \frac{dv}{1-\rho^2 u v} \] where $\rho = M/m = 2 + \Delta/m$
and one would attempt to do some sort of expansion near $\rho =2$. 
Unfortunately such an expansion is impractical as the integral contains a
pole at $\rho=2, u=1/2$. 

\section{Results}
\subsection{$K^*\to K \gamma$ and the coupling ratio}
The observed $K^*$ branching fraction of 
\[ 
 \decay{K^{*0}}{K^0 \gamma} / \decay{K^{*+}}{K^+ \gamma} = 2.31 \pm 0.29 
\]
and corresponding coupling constant ratio of
\begin{equation}
 \mod{ g_{K^{*0} K^0 \gamma}/g_{K^{*+} K^+ \gamma}}  = 1.514 \pm 0.125,
\label{eq:Kcr}
\end{equation}
is far from its SU(3) predicted value of 2, but
is simply understood in the quark loop formalism as shown by 
Bramon and Scadron \cite{bram8912}.  We quickly re--iterate 
this point.  Using Eq.\ (\ref{eq:couple}) and assuming $g_{Vus}=
g_{Vds}$ and $g_{Pus}=g_{Pds}$ then
\begin{equation}
 \frac{ g_{K^{*0} K^0 \gamma}}{g_{K^{*+} K^+ \gamma}} = 
-\,\frac{J_{d,s}[K^{*0},K^0]/m_d + J_{s,d}[K^{*0},K^0]/m_s}
 {2 J_{u,s}[K^{*+},K^+]/m_u - J_{s,u}[K^{*+},K^+]/m_s} = -1.47,
\label{eq:Kratio}
\end{equation}
where we have used a more complete notation, $J_{q,\bar{q}'}[V,P]$ to
denote $J$ for quarks $q$, $\bar{q}'$, vector meson $V$, and pseudoscalar
meson $P$ of masses $m, \mb, M^*$ and $M$ respectively. We used quark
masses $m_u=m_d=340$ MeV and $m_s=510$ MeV. The experimental uncertainty
in the coupling ratio permits the $s$ quark mass range: $475<m_s<545$ MeV,
when $m_u=340$ MeV, and a fixed $s$ quark mass of $m_s=510$ MeV permits a
$u$ quark range of $225 < m_u < 385$ MeV.  Since an $s$ quark mass of $m_s
= m_{\phi}/2$ gives such a good comparison between the quark triangle
diagram and the experimental measurement, we will choose such a mass
throughout this work, along with $m_u=m_d=340$ MeV.  The result of Eq.\
(\ref{eq:Kratio}) compares well with that of \cite{bram8912}, indicating
their chiral limit formulae is appropriate.  In fact we can observe the
variation from the chiral limit to $M = M_K$ using our $J$; as
Figure~\ref{fig:chiralK} shows there is very little change in the result. 

We can also dramatically show how the $s$ quark mass breaks the SU(3)
symmetry.  Figure~\ref{fig:m_s} displays the behaviour of $ g_{K^{*0} K^0
\gamma}/g_{K^{*+} K^+ \gamma}$ from $m_s=m_u=m_d=340$ MeV (the SU(3) limit) to
$m_s=550$ MeV clearly indicating that it is the violation of constituent
quark masses from SU(3) symmetry that is responsible for the large
deviation in $K^*$ mesons from the expected symmetry.  The $K^*$ radiative
decays are particularly sensitive to SU(3) violations as they involve the
constituent masses of strange and non--strange quarks in the loop of the
corresponding triangle diagram.  Subsequently, the heavier meson cases are
best computed by the triangle diagram which can allow for different
constituent masses in the loop, rather than an SU(4) or SU(5) symmetry.

\subsection{Measurements of $\gP$}
We may obtain estimates for the $\gP$ coupling constants using
experimental measurements of the $P \to \gamma\gamma$ decay widths.  In particular
the widths for $\pi^0 \to \gamma\gamma$, $\eta \to \gamma\gamma$ and $\eta \prime \to
\gamma\gamma$ processes are well known and we should be able to determine $\gPu$
and $\gPs$ to reasonable accuracy (we make the isosymmetric approximation
$\gPu = g_{Pd\bar{d}}, \; m_u = m_d$). 

Beginning with the $\pi^0$ meson which is the antisymmetric mixture 
$(u\bar{u} - d\bar{d})/\sqrt{2}$, we find
\begin{eqnarray}
 g_{\pi^0\gamma\gamma} &=& \frac{6 e^2}{4 \pi^2} \left\{ \left(\frac{2}{3} \right)^2
	g_{\pi^0 u \bar{u}} \frac{J_{u}[\pi^0]}{m_u} + 
\left(- \frac{1}{3} \right)^2 g_{\pi^0 d\bar{d}} \frac{J_{d}[\pi^0]}{m_d} 
	\right\} \nonumber \\ 
	&=&  \frac{e^2}{6 \pi^2} \left\{ 4 \frac{\gPu}{\sqrt{2}} - 
	\frac{g_{Pd\bar{d}}}{\sqrt{2}} \right\} J_u[\pi^0]/m_u \nonumber \\
	&=& \frac{e^2}{2 \sqrt{2} \pi^2} \gPu J_u[\pi^0]/m_u. 
\label{eq:P1}
\end{eqnarray}
In a similar fashion, we use the standard octet-singlet pseudoscalar 
mixing angle $\tP$ to ascribe the $\eta$--$\eta\prime$ mixing by
\begin{eqnarray*}
\eta &=& \frac{1}{\sqrt{6}}[ (\cos \tP - \sqrt{2} \sin \tP)
(u\bar{u} + d\bar{d}) - \sqrt{2} (\sin\tP+ \sqrt{2} \cos\tP) s\bar{s}], \\
\eta\prime &=& \frac{1}{\sqrt{6}}[ (\sin \tP + \sqrt{2}\cos \tP)
(u\bar{u} + d\bar{d}) + \sqrt{2}(\cos\tP - \sqrt{2}\sin\tP) s\bar{s}]. 
\end{eqnarray*}
Following a methodology like Eq.\ (\ref{eq:P1}) we obtain relations
between the covariant amplitudes and meson--quark couplings which are
given in Table \ref{tab:gVgP}.  We have used $m_u =340$ MeV, $m_s=510$
MeV, two different mixing angles, $\tP=\dg{-10.5}$ (in accordance with the
quadratic Gell-Mann--Okubo relation) and $\tP=\dg{-20}$ (which is partly
favored by pseudoscalar decay processes) along with the covariant
couplings from the measured decay rates \cite{pdg94} 
\[ \decay{P}{\gamma\gamma} = m^3_P g_{P\gamma\gamma}^2 / 64 \pi, \] 
to obtain several estimates of the pseudoscalar--quark couplings as given
in Table~\ref{tab:results}.  We used the $\eta \to \gamma\gamma$ and $\eta\prime
\to \gamma\gamma$ to determine simultaneously the values. 

From the results, it appears $\gPu$ differs as determined from $\pi^0$,
$\eta$ and $\eta\prime$ processes. Considering that the $\eta$ meson is
about four times as massive as the pion, it may be appropriate to allow
for such a mass dependency in the coupling constant.  Suppose we label the
first coupling constant from $\pi^0 \to \gamma\gamma$ as $\gPu (m_{\pi^0}^2)$,
while the second from $\eta \to \gamma\gamma$ and $\eta\prime \to \gamma\gamma$ as a
coupling constant somewhere between $m_{\eta}$ and $m_{\eta\prime}$. 
Numerically we took the appropriate mass as the equal weight average,
$(m_{\eta}^2+ m_{\eta\prime}^2)/2$. By linearly interpolating between
these two couplings, we estimated a value of $\gPu (m_{\eta}^2) = 4.61 \pm
0.19$ for $\tP=\dg{-10.5}$ and $\gPu (m_{\eta}^2) = 4.39 \pm 0.17$ for
$\tP = \dg{-20}$ . 

The Goldberger--Treiman (GT) relation at the quark--level, gives us a good
check of our results.  For the pion, the relation reads
\[      f_{\pi} \gPu(m_{\pi^0}^2) / \sqrt{2} = m_u.  \] 
Using our coupling value in Table~\ref{tab:results} along with $m_u = 340$
MeV we predict $f_{\pi} = 93.5 \pm 3.46$ MeV which compares well with the
experimental result $f_{\pi} = 92.4 \pm 0.26$ MeV \cite{pdg94}. 

Also included in Table~\ref{tab:results} is the estimate of $\gPc$ using a
charm quark mass of $m_c = 1550$ MeV along with the experimentally
determined width \cite{pdg94} of $\decay{\eta_c}{\gamma\gamma} = 7.0 \pm 2.6$ keV. 

\subsection{Measurements of $\gV$}
There exist many useful decay channels $V \to P\gamma$ and corresponding data
from which we can determine the product $\gV \gP$. To this end we proceed
in two steps.  Firstly, we interpret individual meson--meson--photon
couplings in terms of meson--quark--antiquark couplings, deriving
relations between them as shown in Table~\ref{tab:gVgP}. Assuming isospin
symmetry there are only two unknown products of couplings involved in the
light meson sector; one is $\gVu \gPu$ for non--strange quarks while the
other is $\gVs\gPs$ for strange quarks. 

Following this we extract individual meson--meson--photon couplings from
the most recently measured decay widths $\decay{V}{P\gamma} = (m_V^2-m_P^2)^3
\gvpg^2/12 \pi m_V^3$ by simply removing the kinematic factors.  The
results are listed in the first column of Table~\ref{tab:results}.  As one
can see, they scatter over a relatively wide range. 

We are able to determine $\gVu\gPu$ solely from any one of the processes
$\rho^0 \to \pi^0 \gamma,\; \rho^+ \to \pi^+ \gamma, \; \rho^0 \to \eta \gamma, \;
\omega \to \pi^0 \gamma$ and $\omega \to \eta \gamma$.  In addition, the decays
$\omega \to \eta \gamma$ and $\phi \to \eta \gamma$ can be used to simultaneously
solve for $\gVu\gPu$ and $\gVs\gPs$.  Our numerical results are shown in
the second column of Table~\ref{tab:results} where we use the same quark
masses as previously along with standard mixing angles. 

The values of the product $\gVu\gPu$ turn out to lie in a quite small
range, except for that from the $\phi \to \pi^0\gamma$.  However, it would
fall into this range had we chosen a mixing angle of about $\tV =
\dg{224}$, a change of $\dg{4.6}$.  Such a high sensitivity of $\phi \to
\pi^0\gamma$ to change in mixing angle suggests it is reasonable to exclude
this channel from our analysis.  Recalling the couplings of pseudoscalar
meson with quark--antiquark pairs discussed previously, we now obtain
$\gVu$. 

As the light vector meson masses vary by less than $30\%$, we shall not
attempt to distinguish between the slightly different couplings and thus
on average we find $\gVu = 2.40 \pm 0.08$ (weighted average) for $\tP =
\dg{-10.5}$ and $\gVu = 2.35 \pm 0.08$ (weighted average) for $\tP =
\dg{-20}$.  It differs from $\gPu$, revealing a substantial violation of
the spin symmetry in the triangle scheme. 

We repeat this procedure in the analysis of $\gVs$, but with fewer
channels to determine a result.  Consequently we have $\gVs = 1.10$ for
$\tP = \dg{-10.5}$ and $\gVs = 1.06$ for $\tP = \dg{-20}$, indicating a
large SU$(3)_V$ symmetry breaking once again. Estimates for $\gVc$ using
the $J/\psi \to \eta_c \gamma$ channel yield $\gVc = 0.92 \pm 0.23$.  Note
that $\gVc$ and $\gPc$ are not substantially different, perhaps indicative
of a limit $g_{Vq\bar{q}} = g_{Pq\bar{q}}$ as $m_q$ gets large. 

For completeness, we wish to obtain a measure of $g_{Vu\bar{s}}$ using the
product $g_{Vu\bar{s}}g_{Pu\bar{s}}$.  However, we have no means of
getting $g_{Pd\bar{s}}$ for the kaon in the triangle scheme.  This is
because unlike $\pi^0$, the $K^0 \to \gamma\gamma$ decay is not mediated by pure
electromagnetic interactions.  However assuming the Goldberger--Treiman
relation at the quark level
\[      f_K g_{Pd\bar{s}}(m_K^2) = (m_u+m_s)/2  \]
we find $g_{Pd\bar{s}} = 3.77 \pm 0.03$ where we have used $f_K = 113.0
\pm 1.0$ MeV \cite{pdg94}.  Subsequently, $g_{Vd\bar{s}} = 2.21 \pm 0.10$
(averaged over the charged and neutral processes). 

We ought to point out that the triangle scheme has also been applied to
radiative decays of $\eta\prime$ into $\rho^0$ or $\omega$, but failed to
yield coupling constants near the above range. This suggests to us that we
should treat $\eta\prime$ in a different way which would most likely
incorporate the U(1) anomaly. 

\section{Predictions}
\subsection{$\phi \to \eta\prime \gamma$ coupling constant and 
branching fraction}
We can use our best fit estimates of the meson--quark coupling
constants to predict the decay width for the decay $\phi \to \eta\prime
\gamma$.  
\begin{eqnarray*} g_{\phi \to \eta\prime \gamma} &=& \frac{e}{12 \pi^2} \{
(\cos\tV -\sqrt{2}\sin\tV)(\sin\tP+\sqrt{2}\cos\tP) \gVu\gPu 
J_{u,u}[\phi,\eta\prime]/m_u +  \\
 && \hspace*{1cm} 2(\sin\tV+\sqrt{2}\cos\tV) (\cos\tP-\sqrt{2}\sin\tP) 
\gVs \gPs J_{s,s}[\phi,\eta\prime]/m_s \} \\
\end{eqnarray*}
and using $m_u =340$ MeV, $m_s = 510$ MeV and mixing angles $\tP = \dg{-
10.5}, \tV = \dg{219.4}$ along with our couplings from
Table~\ref{tab:results}, we compute the coupling to be 
$g_{\phi \to \eta\prime \gamma} = -5.69 \times 10^{-4}$. This gives a 
branching ratio of
\[ Br(\phi \to \eta\prime \gamma) = 4.16 \times 10^{-4} \]
which is slightly above the experimental upper limit of $ Br(\phi \to
\eta\prime \gamma) < 4.1 \times 10^{-4}$ at 90\% confidence level
\cite{pdg94}. However, we note that a change to an $s$ quark mass of $m_s
= 500$ MeV produces a branching fraction of $Br(\phi \to \eta\prime \gamma) =
3.24 \times 10^{-4}$, so that the result displays very sensitive
dependence on the choice of $s$ quark mass, and probably vector mixing
angle.  Also we remain cautious of predictions involving the $\eta\prime$
meson due to its association with the U(1) anomaly. 

\subsection{$D^* \to D\gamma$ and $B^* \to B\gamma$ coupling ratios} 
Since our $J$ is approximation free (i.e.\ no chiral limit assumptions)
we can safely use it in the $D^*$ and 
$B^*$ meson cases.  We do assume $g_{Vu\bar{Q}}=g_{Vd\bar{Q}}$ and
$g_{Pu\bar{Q}} = g_{Pd\bar{Q}}$ where $Q$ is either the $c$ or $b$
quark (much like we did in the $K^* \to K\gamma$ case) to obtain
\begin{equation} 
\frac{ g_{D^{*0} D^0 \gamma}}{g_{D^{*+} D^+ \gamma}} = 
\,\frac{2 ( J_{u,c}[D^{*0},D^0]/m_u + J_{c,u}[D^{*0},D^0]/m_c)}
 {- J_{d,c}[D^{*+},D^+]/m_d + 2 J_{c,d}[D^{*+},D^+]/m_c},
\label{eq:Dratio}
\end{equation}
and
\begin{equation} 
\frac{ g_{B^{*0} B^0 \gamma}}{g_{B^{*+} B^+ \gamma}} = 
\,\frac{J_{d,b}[B^{*0},B^0]/m_d + J_{b,d}[B^{*0},B^0)/m_b]}
 { J_{u,b}[B^{*+},B^+]/m_u - 2 J_{b,u}[B^{*+},B^+]/m_b}.
\label{eq:Bratio}
\end{equation}
Relations (\ref{eq:Dratio}) and (\ref{eq:Bratio}) allow us examine the
coupling constant ratios as a function of the $c$ and $b$ quark mass,
respectively.  The results appear in Figures~\ref{fig:Dratio} and
\ref{fig:Bratio}.  In order to give actual values we use a $c$ quark mass
of 1550 MeV (approximately half the $J/\psi$ mass) yielding $g_{D^{*0} D^0
\gamma}/g_{D^{*+} D^+ \gamma} = 6.47$ and a $b$ quark mass of 4730 MeV
(approximately half the $\Upsilon$ mass) which gives $g_{B^{*0} B^0
\gamma}/g_{B^{*+} B^+ \gamma} = 0.018$.  We can compare our results with those of
other workers.  These are presented in Table~\ref{tab:comparison}. We
hoped that our study of $\gV, \gP$ measurements would enable us to make
some reasonable guesses of $g_{Vu\bar{c}} g_{Pu\bar{c}}$ and
$g_{Vu\bar{b}} g_{Pu\bar{b}}$, but the data does not allow this.  Thus we
cannot make predictions about actual decay widths. 

\section{Conclusions}
We have successfully evaluated $V \to P \gamma$ and $P\to\gamma\gamma$ processes in a
quark triangle diagram scheme which is valid for arbitrary vector or
pseudoscalar masses.  By comparison with available experimental data, we
found that this scheme works well for all radiative processes involving
the light mesons (no charm or bottom quarks), except for $\phi \to \pi^0
\gamma$ (due to the sensitivity of this channel to the mixing angle) and
$\eta\prime \to \rho^0 (\omega) \gamma$. 

The scheme produces well determined estimates of the
meson--quark--antiquark couplings for the light mesons.  The large
difference between $\gV$ and $\gP$ indicates a substantial violation of
spin symmetry in the quark triangle formalism. We also observed a
relatively weak SU(3) chiral symmetry breaking due to the finite masses of
the Goldstone--type pseudoscalar mesons, along with a more apparent
SU$(3)_V$ symmetry breakdown arising from the difference in light
constituent quark masses.  We note that these coupling constants are
relatively insensitive to the pseudoscalar mixing angle. 

A number of predictions have been made based on the scheme.  Firstly we
note that our theoretical result for the $\phi \to \eta\prime \gamma$ decay
width is around the present experimental upper limit and awaits comparison
with further measurement.  Secondly our prediction for $g_{D^{*0} D^0\gamma}/
g_{D^{*\pm} D^{\pm}\gamma} = 6.47$ with $m_c \approx m_{J/\psi}/2$ is within
range of other theoretical estimates, while $g_{B^{*0} B^0\gamma}/ g_{B^{*\pm}
B^{\pm}\gamma} = 0.018$ for $m_b \approx m_{\Upsilon}/2$ is small compared
with the few results in the literature.  We expect future measurements of
these radiative decays will distinguish between these predictions. 

\acknowledgements
The authors wish to thank Prof.\ R.\ Delbourgo and Dr.\ D.\ Kreimer for
useful discussions.  Dongsheng Liu wishes to thank the ARC for financial
assistance under Grant Number A69231484 and the organizers of the Joint
Japan Australia Workshop on ``Quarks, Hadrons and Nuclei'' in which Prof.\
M.\ Oka \cite{taki9506} and Dr.\ H.\ Yabu \cite{cheo9503} brought their
papers to his attention.



\setlength{\unitlength}{1cm}

\begin{table}
\[
\begin{array}{c|l}
\mbox{Process} &  \mbox{Relation between covariant couplings 
			and meson--quark couplings} \\ \hline
\pi^0 \to \gamma\gamma & g_{\pi^0\gamma\gamma} = \frac{e^2}{2 \sqrt{2} \pi^2} \gPu
J_u[\pi^0]/m_u \\
\eta \to \gamma\gamma & g_{\eta\gamma\gamma} = \frac{e^2}{6 \sqrt{6} \pi^2} 
\{ 5(\cos\tP - \sqrt{2} \sin\tP) \gPu J_{u}[\eta]/m_u - \\ 
 & \hspace*{2cm} \sqrt{2}(\sin\tP +\sqrt{2}\cos\tP) \gPs
J_{s}[\eta]/m_s \}
\\
\eta\prime \to \gamma\gamma & g_{\eta\prime\gamma\gamma} = \frac{e^2}{6 \sqrt{6} \pi^2} 
\{ 5(\sin\tP + \sqrt{2} \cos\tP) \gPu J_{u}[\eta\prime]/m_u + \\ 
 & \hspace*{2cm} \sqrt{2}(\cos\tP -\sqrt{2}\sin\tP) \gPs
J_{s}[\eta\prime]/m_s \} \\
\eta_c \to \gamma\gamma & g_{\eta_c \to \gamma\gamma} = \frac{2 e^2}{3 \pi^2} 
		\gPc J_c[\eta_c]/m_c\\
\rho^0 \to \pi^0 \gamma & g_{\rho^0 \to \pi^0 \gamma} = \frac{e}{4 \pi^2} \gVu \gPu
   J_{u,u}[\rho^0,\pi^0]/m_u \\
\rho^+ \to \pi^+ \gamma & g_{\rho^+ \to \pi^+ \gamma} = \frac{e}{4 \pi^2} 
g_{Vu\bar{d}} g_{Pu\bar{d}} J_{u,d}[\rho^+,\pi^+]/m_u \\
\rho^0 \to \eta \gamma & g_{\rho^0 \to \eta \gamma} = \frac{\sqrt{3}e}{4 \pi^2} 
(\cos\tP - \sqrt{2} \sin\tP) \gVu \gPu 
   J_{u,u}[\rho^0,\eta]/m_u  \\
\omega \to \pi^0 \gamma & g_{\omega \to \pi^0 \gamma} = \frac{\sqrt{3} e}{4 \pi^2} 
(\sin\tV+ \sqrt{2} \cos\tV ) \gVu \gPu J_{u,u}[\omega,\pi^0]/m_u \\
\omega \to \eta \gamma & g_{\omega \to \eta \gamma} = \frac{e}{12 \pi^2} \{ \\
& \hspace*{1cm}
(\sin\tV+\sqrt{2}\cos\tV)(\cos\tP-\sqrt{2}\sin\tP) \gVu\gPu 
J_{u,u}[\omega,\eta]/m_u +  \\
 & \hspace*{1cm} 2(\cos\tV-\sqrt{2}\sin\tV) (\sin\tP + \sqrt{2} \cos\tP) 
\gVs \gPs J_{s,s}[\omega,\eta]/m_s \} \\
\eta\prime \to \rho^0 \gamma & g_{\eta\prime \to \rho^0 \gamma} = 
 \frac{\sqrt{3} e}{4 \pi^2} 
(\sin\tP+\sqrt{2}\cos\tP) \gVu \gPu J_{u,u}[\eta\prime,\rho^0]/m_u  \\
\eta\prime \to \omega \gamma & g_{\eta\prime \to \omega \gamma} = 
\frac{e}{12 \pi^2} \{ \\
 & \hspace*{1cm} (\sin\tV+\sqrt{2}\cos\tV)(\sin\tP+\sqrt{2}\cos\tP)
 \gVu \gPu J_{u,u}[\eta\prime,\omega]/m_u -  \\
 & \hspace*{1cm} 2 (\cos\tV -\sqrt{2}\sin\tV)(\cos\tP -\sqrt{2}\sin\tP)
 \gVs \gPs J_{s,s}[\eta\prime,\omega]/m_s \} \\
\phi \to \pi^0 \gamma & g_{\phi \to \pi^0 \gamma} = \frac{\sqrt{3} e}{4 \pi^2} 
(\cos\tV - \sqrt{2} \sin\tV) \gVu \gPu J_{u,u}[\phi,\pi^0]/m_u \\
\phi \to \eta \gamma & g_{\phi \to \eta \gamma} = \frac{e}{12 \pi^2} \{ \\
& \hspace*{1cm}
(\cos\tV -\sqrt{2}\sin\tV)(\cos\tP-\sqrt{2}\sin\tP) \gVu\gPu 
J_{u,u}[\phi,\eta]/m_u -  \\
 & \hspace*{1cm} 2(\sin\tV+\sqrt{2}\cos\tV) (\sin\tP + \sqrt{2} \cos\tP)
\gVs \gPs J_{s,s}[\phi,\eta]/m_s \} \\
K^{*0} \to K^0\gamma & g_{K^{*0} \to K^0\gamma} = \frac{e}{4\pi^2} g_{Vd\bar{s}} 
g_{Pd\bar{s}} ( J_{d,s}[K^{*0},K^0]/m_d + J_{s,d}[K^{*0},K^0]/m_s) \\
K^{*+} \to K^+\gamma & g_{K^{*+} \to K^+\gamma} = \frac{e}{4\pi^2} g_{Vu\bar{s}}
 g_{Pu\bar{s}} ( 2 J_{u,s}[K^{*+},K^+]/m_u - J_{s,u}[K^{*+},K^+]/m_s) \\
J/\psi \to \eta_c\gamma & g_{J/\psi \to \eta_c\gamma} = \frac{e}{\pi^2} \gVc 
\gPc J_{c,c}[J/\psi,\eta_c]/m_c \\
\hline
\end{array}
\]
\caption{Radiative decays of ground state mesons and relations 
between covariant couplings $g_{P\gamma\gamma}, g_{VP\gamma}$ or
$g_{PV\gamma}$ and meson--quark--antiquark couplings $\gV,\gP$}
\label{tab:gVgP}
\end{table}

\begin{table}
\[
\begin{array}{clr}
\mbox{Experimental result} &
\multicolumn{2}{c}{\mbox{Meson--quark--antiquark coupling}\tablenote{
 $m_u=m_d=340$ MeV, $m_s=510$ MeV, $m_c=1550$ MeV, $\tV=\dg{219.4}$}} \\      
(\times 10^{-4} \mbox{MeV}^{-1}) \mbox{\cite{pdg94}}  
	& \tP = \dg{-10.5} & \tP= \dg{-20} \\ \hline 
\mod{g_{\pi^0\gamma\gamma}} = 0.2516 \pm 0.0091 & \multicolumn{2}{c}{
	\gPu = 5.14 \pm 0.19} \\
\left. \begin{array}{c}
\mod{g_{\eta\gamma\gamma}} = 0.239 \pm 0.011 \\
\mod{g_{\eta\prime\gamma\gamma}} = 0.312\pm 0.016 
\end{array} \right\}  & 
\left\{ \begin{array}{l}
  \gPu= 4.03 \pm 0.14, \\ \gPs = 6.42 \pm 0.67 
\end{array} \right. &
\left\{ \begin{array}{r}
  \gPu= 3.56 \pm 0.12, \\ \gPs = 8.19 \pm 0.65 
\end{array} \right.
\\
\mod{g_{\eta_c\gamma\gamma}} = 0.07297 \pm 0.01366 & 
	\multicolumn{2}{c}{\gPc = 2.03 \pm 0.38}\\
\mod{g_{\rho^0 \to \pi^0 \gamma}} = 2.96 \pm 0.38 & 
	\multicolumn{2}{c}{\gVu \gPu =  14.82 \pm 1.90} \\
\mod{g_{\rho^+ \to \pi^+ \gamma}} = 2.24 \pm 0.13 & 
	\multicolumn{2}{c}{\gVu \gPu = 11.32 \pm 0.64} \\
\mod{g_{\rho^0 \to \eta \gamma}} = 5.67 \pm 0.53 & \gVu \gPu = 10.96 \pm 1.02 &
	\gVu \gPu = 9.56 \pm 0.89 \\
\mod{g_{\omega \to \pi^0 \gamma}} = 7.04 \pm 0.21 & 
	\multicolumn{2}{c}{\gVu \gPu = 12.53 \pm 0.38} \\
\left. \begin{array}{c}
\mod{g_{\omega \to \eta \gamma}} = 1.83 \pm 0.23  \\
\mod{g_{\phi \to \eta \gamma}} = 2.117 \pm 0.052 
\end{array} \right\} & 
\left\{ \begin{array}{l} 
 \gVu\gPu = 12.3\pm 1.5,\\  \gVs \gPs = 7.08 \pm 0.17
\end{array} \right. &
\left\{ \begin{array}{r} 
 \gVu\gPu = 10.7 \pm 1.3,\\  \gVs \gPs = 8.66 \pm 0.21
\end{array} \right. \\
\mod{g_{\phi \to \pi^0 \gamma}} = 0.417 \pm 0.021 & 
	\multicolumn{2}{c}{\gVu\gPu = 25.1 \pm 1.3
	\tablenote{$\gVu\gPu = 11.9 \pm 0.6$ for $\tV=\dg{224}$}}\\
\mod{g_{K^{*0} \to K^0\gamma}} = 3.84 \pm 0.17 & 
	\multicolumn{2}{c}{g_{Vd\bar{s}} g_{Pd\bar{s}} = 8.43 \pm 0.37} \\
\mod{g_{K^{*+} \to K^+\gamma}} = 2.534 \pm 0.115 & 
	\multicolumn{2}{c}{g_{Vu\bar{s}} g_{Pu\bar{s}} = 8.21 \pm 0.37}\\
\mod{g_{J/\psi \to \eta_c\gamma}} = 1.67 \pm 0.26 & 
	\multicolumn{2}{c}{\gVc\gPc = 1.87 \pm 0.30} \\
\hline
\end{array} 
\]
\caption{Determination of meson--quark--antiquark couplings}
\label{tab:results}
\end{table}

\begin{table}
\begin{tabular}{ccc}
Reference   & $ \mod{g_{D^{*0} D^0 \gamma}/ g_{D^{*+} D^+ \gamma}} $
	    & $\mod{g_{B^{*0} B^0 \gamma}/g_{B^{*+} B^+ \gamma}}$
\\  \hline
\cite{pdg94} & - & - \\
This paper & 6.47 & 0.018 \\
\cite{alie9511} & $3.05 \pm 0.63$ & $0.49 \pm 0.38$ \\
\cite{suth95} & $2.98 \pm 0.62$ & - \\
\cite{dosc95} & $6.32 \pm 2.97$ & $0.64\pm  0.51$ \\
\cite{chen}\tablenote[1]
	{Gaussian wave function with $m_u$=300MeV, $m_c$=1500MeV}
	& $11.0\pm 2.3$ & - \\
\cite{chen}\tablenote[2]
	{BS wave function with $m_u$=350MeV, $m_c$=1500MeV}
	& $12.9\pm 2.7$ &  - \\
\cite{alie94} & $3.28\pm0.83$ & - \\
\cite{fayy94} & $5.4 \pm 1.1$ & - \\
\cite{jain9407} & - & $0.58\pm 0.44$  \\
\cite{odon9407}
	\tablenote[3]{$m_u=m_d=300$ MeV, $m_s=500$ MeV} 
	& $6.95\pm 1.44$ & - \\
\cite{cola9407} & $6.61\pm 1.37$ & $0.62 \pm 0.48$ \\
\cite{cola9307} &$5.54\pm 3.00$ & $0.59 \pm 0.48$ \\  
\cite{jones94} & $60$ & $0.8$
	\tablenote[4]{The original paper contained an error for
	this calculation.  This is the corrected value.} \\
\cite{kama} & $3.50\pm 0.73$& - \\
\cite{sing} & - & $0.678\pm 0.523$ \\
\cite{rosn} & $3.84 \pm 0.80$ & - \\
\cite{mill}
	\tablenote[5]{Zero anomalous magnetic moment of charm quark} 
	& $1.66\pm 0.34$ & - \\
\cite{thew}
	\tablenote[6]{SU(4) symmetry}
	& $3.93\pm 0.84$ & -\\
\cite{thew}
	\tablenote[7]{Broken SU(4) by $M1$ transition} 
	& $4.49\pm 0.96$ & -\\
\cite{thew}
	\tablenote[8]{Broken SU(4) by $1/M_V^2$}
	& $3.92\pm 0.84$ & -\\
\cite{pham} & $2.00 \pm 0.41$ & - \\
\cite{eich} & $3.78\pm 0.78$ & - \\
\end{tabular}
\caption{Summary of theoretical estimates}
\label{tab:comparison}
\end{table}

\newpage
\begin{figure}
\begin{picture}(16,12)
\put (0.5,-4.5){\epsfxsize=13cm \epsfbox{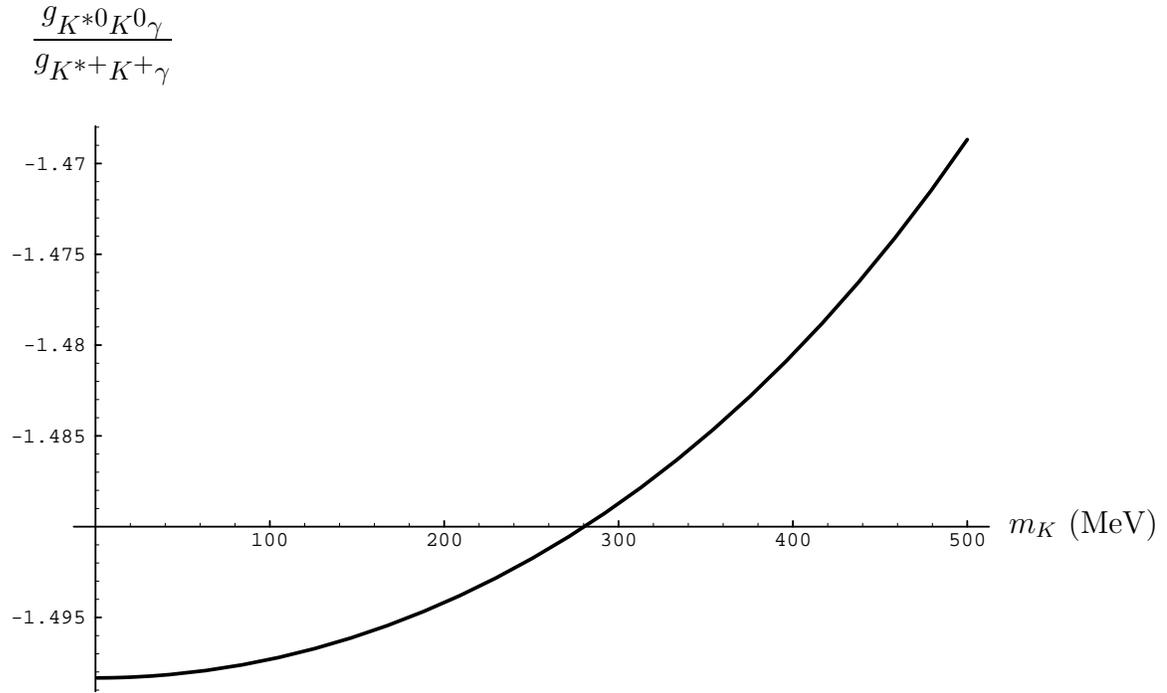}}
\put (13.75,2.85){$m_K$ (MeV)}
\put (0.75,9.25){\Large{$\frac{ g_{K^{*0} K^0 \gamma}}{g_{K^{*+} K^+ \gamma}}$} }
\end{picture}
\caption{Appropriateness of chiral limit, shown in the lack of 
sensitivity of $g_{K^{*0} K^0 \gamma}/g_{K^{*+} K^+\gamma}$ to $K$ meson mass.}
\label{fig:chiralK}
\end{figure}

\newpage
\begin{figure}
\begin{picture}(16,12)
\put (0.5,-4.5){\epsfxsize=13cm \epsfbox{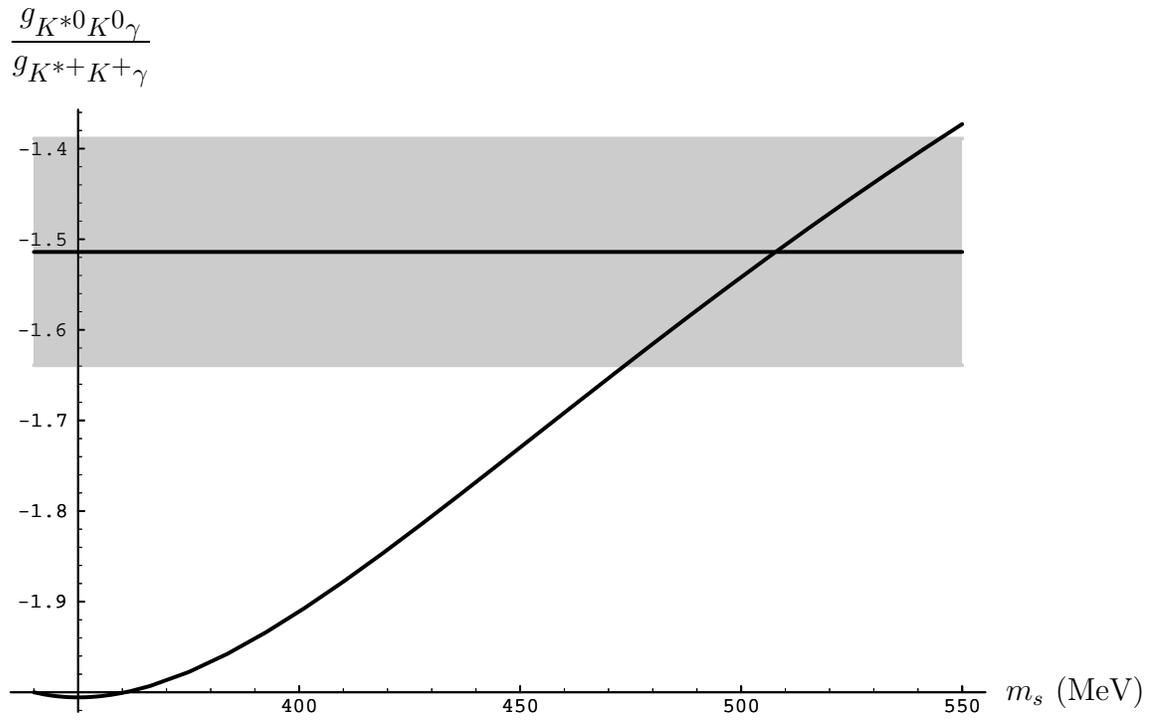}}
\put (13.75,0.65){$m_s$ (MeV)}
\put (0.5,9.25){\Large{$\frac{ g_{K^{*0} K^0 \gamma}}{g_{K^{*+} K^+ \gamma}}$} }
\end{picture}
\caption{Breaking of SU(3) by $s$ quark mass. The experimental measurement is
	included.}
\label{fig:m_s}
\end{figure}

\newpage
\begin{figure}
\begin{picture}(16,12)
\put (0.5,-4.5){\epsfxsize=13cm \epsfbox{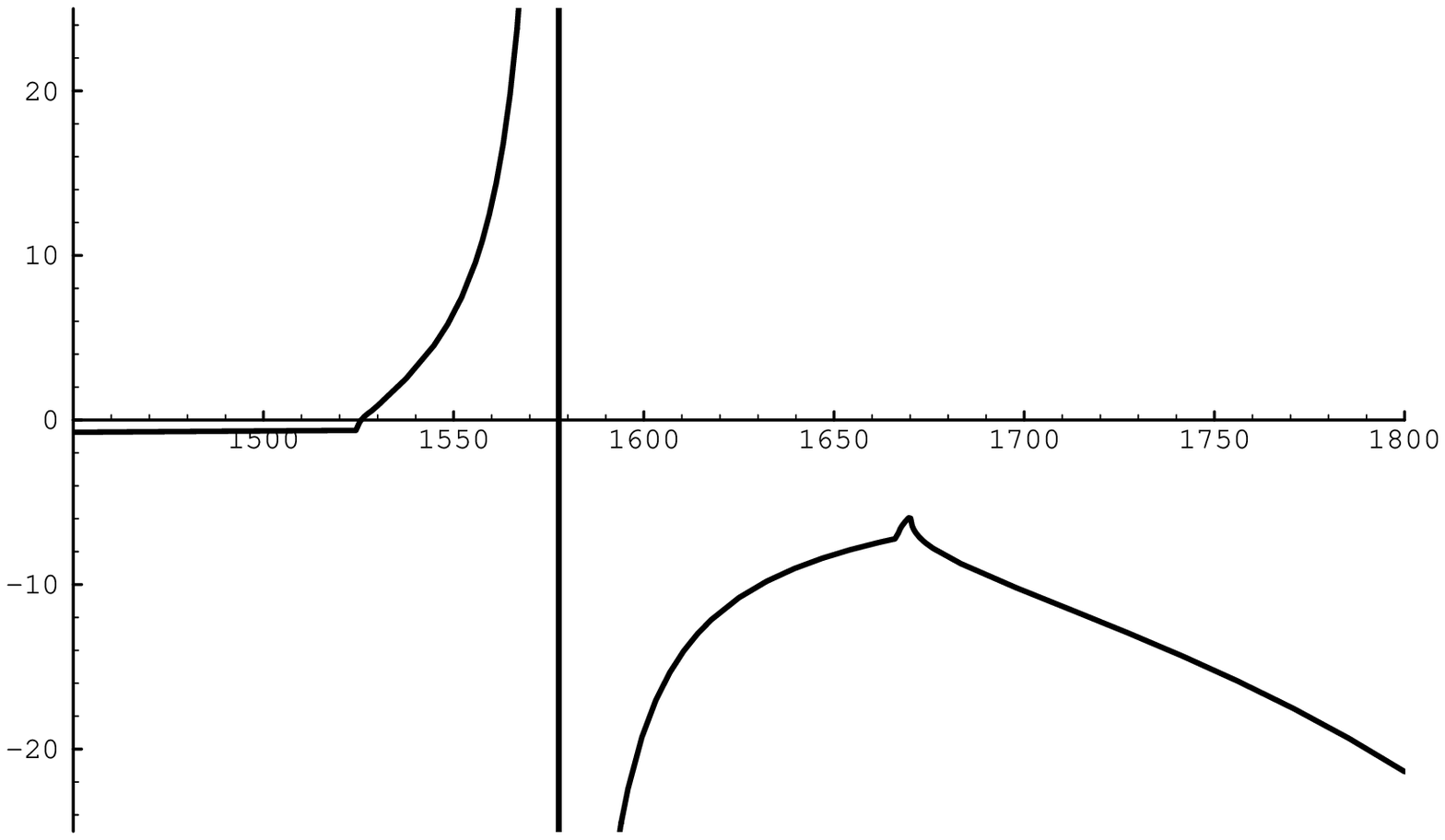}}
\put (13.75,4.4){$m_c$ (MeV)}
\put (0.3,9.25){\Large{$\frac{ g_{D^{*0} D^{0} \gamma}}{g_{D^{*+} D^{+} \gamma}}$} }
\end{picture}
\caption{Variation of coupling ratio with $c$ quark mass.}
\label{fig:Dratio}
\end{figure}

\newpage
\begin{figure}
\begin{picture}(16,12)
\put (0.5,-4.5){\epsfxsize=13cm \epsfbox{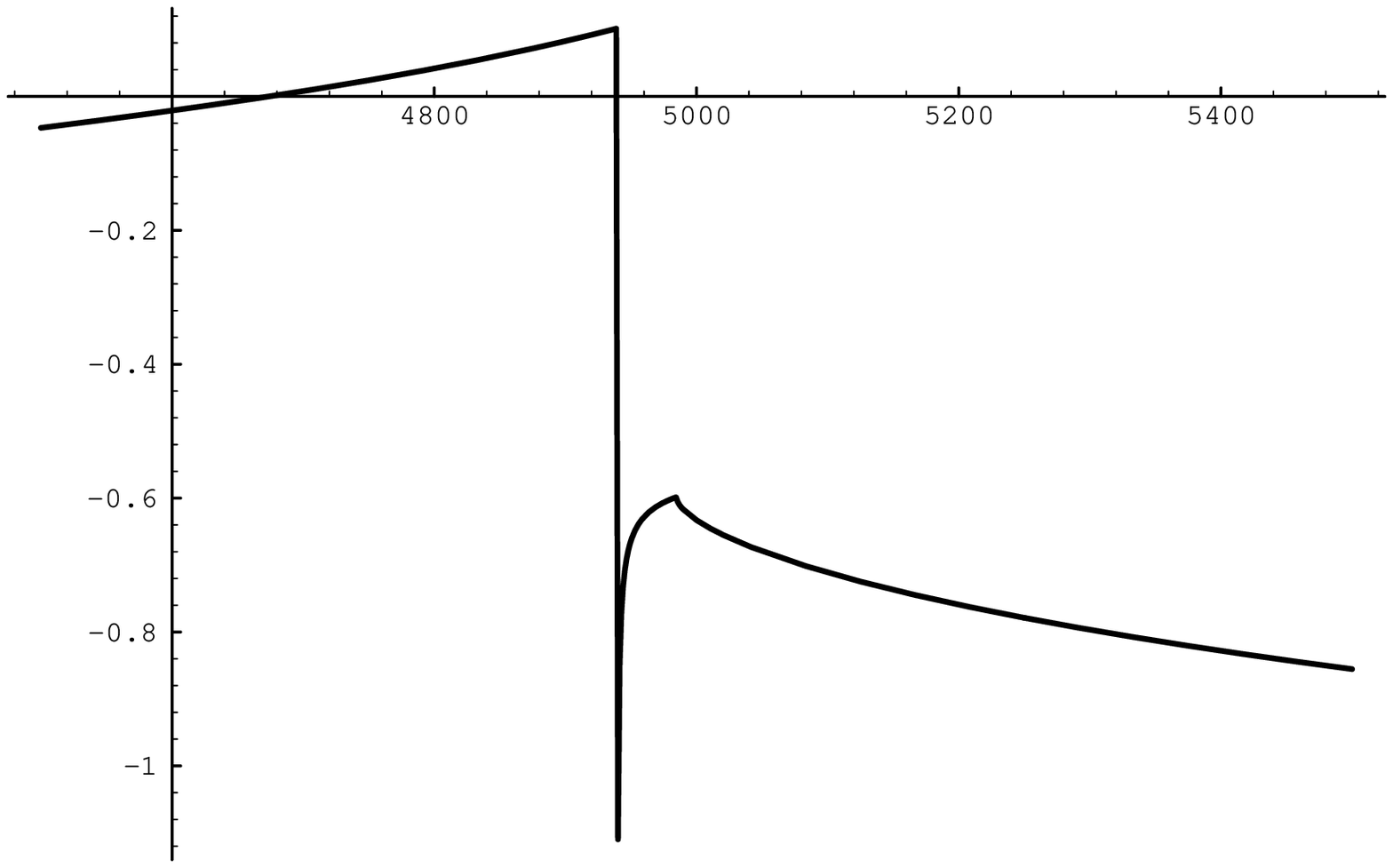}}
\put (13.75,7.55){$m_b$ (MeV)}
\put (1.3,9.25){\Large{$\frac{ g_{B^{*0} B^0 \gamma}}{g_{B^{*+} B^+ \gamma}}$} }
\end{picture}
\caption{Variation of coupling ratio with $b$ quark mass.}
\label{fig:Bratio}
\end{figure}

\end{document}